\documentclass[12pt,a4paper,titlepage]{article}

\usepackage{amssymb}

\advance\textwidth3cm
\advance\textheight4cm
\advance\oddsidemargin-2cm
\advance\topmargin-2.5cm

\def\id{{\mathchoice
{\rm 1 \mskip-4mu l}
{\rm 1 \mskip-4mu l}
{\rm 1 \mskip-4.5mu l}
{\rm 1 \mskip-5mu l}}}

\newcommand{\be}{\begin{equation}}
\newcommand{\ee}{\end{equation}}
\newcommand{\bea}{\begin{eqnarray}}
\newcommand{\eea}{\end{eqnarray}}
\newcommand{\beas}{\begin{eqnarray*}}
\newcommand{\eeas}{\end{eqnarray*}}

\newcommand{\D}{{\sf D}}
\renewcommand{\H}{{\sf H}}
\renewcommand{\L}{{\sf L}}
\newcommand{\M}{{\sf M}}
\newcommand{\N}{{\sf N}}
\renewcommand{\O}{{\sf O}}
\renewcommand{\P}{{\sf P}}
\newcommand{\Q}{{\sf Q}}
\newcommand{\R}{{\sf R}}

\renewcommand{\DH}{\hat{\sf D}}
\newcommand{\HH}{\hat{\sf H}}
\newcommand{\LH}{\hat{\sf L}}

\newcommand{\OH}{\hat{\sf O}}
\newcommand{\PH}{\hat{\sf P}}
\newcommand{\QH}{\hat{\sf Q}}
\newcommand{\RH}{\hat{\sf R}}

\renewcommand{\u}{\vec{u}}
\renewcommand{\v}{\vec{v}}
\newcommand{\p}{\vec{p}}
\newcommand{\pe}{\vec{\pi}}
\newcommand{\x}{\vec{x}}

\newcommand{\e}{\varepsilon}

\newcommand{\C}{{\cal C}}
\newcommand{\HI}{{\cal H}}
\newcommand{\V}{{\cal V}}

\newcommand{\pa}{\partial}

\newcommand{\SL}{\mbox{{\it SL}}(2,\mathbb{R})}

\newcommand{\ds}{\displaystyle}

\renewcommand{\c}{constraint}
\newcommand{\obs}{observable}
\newcommand{\occ}{observable content of the constraints}
\newcommand{\fo}{fundamental observables}
\newcommand{\aoo}{algebra of observables}
\renewcommand{\r}{representation}
\newcommand{\ir}{irreducible representation}
\newcommand{\ie}{{\it i.e.}}

\begin{document}

\begin{titlepage}

\begin{flushright} { THEP 99/3\\Universit\"at Freiburg\\April 1999\\
hep-th/9907056}
\end{flushright}
\vspace{1.5cm}
\begin{center}
{\LARGE\bf An $\SL$ Model of Constrained Systems:\\[3mm]
Algebraic Constraint Quantization}\\[2cm]
{\large Michael Trunk}\\[4mm]
{Universit\"at Freiburg\\
Fakult\"at f\"ur Physik\\
Hermann--Herder--Str.\ 3\\
D--79104 Freiburg\\
Germany\\[5mm]
e-mail: trunk@physik.uni-freiburg.de}\\[3cm]
{\bf Abstract}\\[5mm]
\end{center}

\noindent
A reparametrization invariant model, introduced recently by Montesinos,
Rovelli and Thiemann, possessing an $\SL$ gauge symmetry is treated
along the guidelines of an algebraic \c\ quantization scheme that
translates the vanishing of the \c s into \r\ conditions for the
\aoo. The application of this algebraic scheme to the $\SL$ model
yields an unambiguous identification of the physical \r\ of the
\aoo.

\end{titlepage}

\section{Introduction}

In Ref.\ \cite{MRT}, Montesinos, Rovelli, and Thiemann have introduced
a reparametrization invariant model with an $S\!L(2,\mathbb{R})$
gauge symmetry which, in a finite-dimensional context, ``mimicks
the \c\ structure'' of general relativity.

In this note I do not want to enter into a discussion of the
physical significance of the model. Rather, I will regard the
system from the point of view of an algebraic scheme for the
quantization of constrained systems (presented in Ref.\ \cite{Tr}),
which lays emphasis on the quantization of \obs\ quantities,
translating the ``vanishing'' of the \c s into \r\ conditions
for the \aoo. The $\SL$ model presents an interesting example
for the application of the algebraic scheme, because the model
displays a peculiar structural feature that sheds some light
upon the working of the method (for a
retrospective characterization of this feature, {\it cf}\/ the
conclusions). Also, the treatment of the model presents an
application of the algebraic \c\ quantization scheme to a
reparametrization invariant system with a non-trivial gauge
group (as opposed to the -- almost obvious -- application to
the free relativistic particle), possessing \c s which are
quadratic in the momenta.

The construction of the quantum theory of the $\SL$
model according to the algebraic \c\ quantization scheme results
in an unambiguous identification of the physical \r\ of the \aoo.
Moreover, I show how the construction of the quantum theory
according to the Dirac quantization scheme, that
has been left unfinished in Ref.\ \cite{MRT}, can be completed:
In contrast to the approach followed here, the authors of Ref.\
\cite{MRT} concentrate on the ``quantization'' of the \c s (\ie,
of un\obs\ quantities) which does not automatically yield the
correct quantization of the \obs s.

\section{Classical analysis of the model}

\subsection{Constraints and observables}

The configuration space of the model presented in Ref.\ \cite{MRT}
is $\mathbb{R}^4$, parametrized by the Euclidean variables $\u = (
u_1,u_2)^T$ and $\v=(v_1,v_2)^T$. The model is defined by the
Lagrangian
\[  L = \sqrt{ \v^2 ( \dot{\u} - \lambda \u )^2 } +
        \sqrt{ \u^2 ( \dot{\v} + \lambda \v )^2 } .\]
The ``multiplier'' $\lambda$
can be eliminated with the help of its equation of motion (two
of the multipliers of Ref.\ \cite{MRT} have already been
eliminated), \ie\ by solving the equation $\pa L / \pa \lambda=0$
for $\lambda$: $\lambda = \tilde{\lambda}(\u,\dot{\u},\v,\dot{\v}
)$. The function $\tilde{\lambda}$ being homogeneous of order one
in the velocities, $\tilde{\lambda}(\u,\alpha\dot{\u},\v,\alpha
\dot{\v}) = \alpha \, \tilde{\lambda}(\u,\dot{\u},\v,\dot{\v})$,
the action $S = \int {\rm d}t \: L$ is invariant under
reparametrizations of $t$.

As a consequence of the definition of the canonical momenta,
\[  \p := \frac{\pa L}{\pa \dot{\u}}
        = \left. \frac{\v^2 \, ( \dot{\u} - \lambda \u )}
                {\sqrt{\v^2 ( \dot{\u} - \lambda \u )^2}}
          \right|_{\ds \lambda = \tilde{\lambda}(\u,\dot{\u},\v,\dot{\v})}
         ,\qquad
   \pe := \frac{\pa L}{\pa \dot{\v}}
        = \left. \frac{\u^2 \, ( \dot{\v} + \lambda \v )}
                {\sqrt{\v^2 ( \dot{\v} + \lambda \v )^2}}
          \right|_{\ds \lambda = \tilde{\lambda}(\u,\dot{\u},\v,\dot{\v})},
\]
there are three primary \c s
\beas  \L_1 &=& \frac12 \, ( \H_1 + \H_2 ) = \frac14 \, (\p^{\,2} - \u^2 +
                 \pe^2 - \v^2 ) = 0, \\
       \L_2 &=& \frac12 \, \D = \frac12 \, (\u\cdot\p - \v\cdot\pe) =
                0 , \\
       \L_3 &=& \frac12 \, ( \H_1 - \H_2 ) = \frac14 \, (\p^{\,2} + \u^2 -
                 \pe^2 - \v^2 ) = 0 .
\eeas
The canonical Hamiltonian of the model vanishes, $H = \p\cdot
\dot{\u} + \pe\cdot\dot{\v} - L = 0$. The \c s are first--class,
their Poisson commutation relations close to form a realization
of the Lie algebra $sl(2,\mathbb{R}) = so(2,1) = su(1,1)$:
\[   \{ \L_a,\L_b \} = \e_{ab}^{\;\;\;\: c} \, \L_c ,\qquad\qquad\quad
     \e_{ab}^{\;\;\;\: c} = g^{cd} \e_{abd} , \qquad \e_{123} = 1, \qquad
     a,b,c,d \in \{ 1,2,3 \},
\]
where $g^{ab}$ is the inverse of the $so(2,1)$--metric $g_{ab}
= \mbox{diag}(+,+,-)$, and repeated indices are summed over.

The infinitesimal canonical transformations generated by the
\c s can be integrated to yield an action of the gauge group
$\SL$ on the space $\mathbb{R}^8 = T^\ast \mathbb{R}^4$: it
acts in its defining two-dimensional \r\ on the vectors
\[  \x^{(1)} = \left( \!\!\! \begin{array}{c}  u_1  \\  p_1
               \end{array} \!\!\! \right) ,\qquad
    \x^{(2)} = \left( \!\!\! \begin{array}{c}  u_2  \\  p_2
               \end{array} \!\!\! \right) ,\qquad
    \x^{(3)} = \left( \!\!\! \begin{array}{c}  \pi_1  \\  v_1
               \end{array} \!\!\! \right) ,\qquad
    \x^{(4)} = \left( \!\!\! \begin{array}{c}  \pi_2  \\  v_2
               \end{array} \!\!\! \right),  \]
\ie, $(\x^{(1)},\dots,\x^{(4)}) \stackrel{g \in SL(2,\mathbb{R})}
{\longrightarrow} (g \x^{(1)},\dots,g \x^{(4)})$.

\subsection{Regularization}

The map $\L : \mathbb{R}^8 \to so^*(2,1)$, $(\u,\v,\p,\pe) \mapsto
\L_a(\u,\v,\p,\pe)$ ($so^*(2,1)$ is the dual of the Lie algebra $so
(2,1)$; $\L$ is the so-called momentum map), is not free of critical
points where the \c s $\L_a$ cease to be functionally independent. A
point $q \in \mathbb{R}^8$ is a critical point of $\L$ if and only if
the two vectors
\[  \xi^{(1)} = (u_1,u_2,\pi_1,\pi_2)^T ,\qquad\quad
    \xi^{(2)} = (p_1,p_2,v_1,v_2)^T,  \]
are linearly dependent.
In order to avoid the occurrence of ``physical states'' which cannot
be interpreted in terms of \obs s (see below) the
set $\cal S$ of critical points of $\L$, that is stable under the action
of $\SL$, must be removed from $\mathbb{R}^8$. Consequently,
the phase space of the system is ${\cal P} = (\mathbb{R}^8 \setminus
{\cal S},\bar{\omega})$, with the symplectic form $\bar{\omega}$ that is
induced from the canonical symplectic form $\omega$ on $T^\ast\mathbb{R}^4$.

\subsection{Observables}

The action of $\SL$ leaves invariant the determinants $\O_{ij} :=
\det(\x^{(i)},\x^{(j)}) = - \O_{ji}$. Therefore, the (continuous)
\fo\ \cite{Tr} of the system can be chosen to be
\[   \M_1 := \O_{23} = u_2 v_1 - p_2 \pi_1 ,\quad
     \M_2 := \O_{31} = - u_1 v_1 + p_1 \pi_1 ,\quad
     \M_3 := \O_{12} = u_1 p_2 - u_2 p_1 ,
\]
\[   \N_1 := \O_{14} = u_1 v_2 - p_1 \pi_2 ,\quad
     \N_2 := \O_{24} = u_2 v_2 - p_2 \pi_2 ,\quad
     \N_3 := \O_{43} = v_1 \pi_2 - v_2 \pi_1 .
\]
The Poisson commutation relations of the \fo\ close to form a
realization of the Lie algebra $so(2,2)$:
\[   \{ \M_a,\M_b \} = \e_{ab}^{\;\;\;\: c} \, \M_c ,\qquad
     \{ \M_a,\N_b \} = \e_{ab}^{\;\;\;\: c} \, \N_c ,\qquad
     \{ \N_a,\N_b \} = \e_{ab}^{\;\;\;\: c} \, \M_c ,
\]
which is isomorphic to $so(2,1) \times so(2,1)$, as can be seen
by defining $\Q_a := \frac12 \, ( \M_a + \N_a ) ,\: \P_a := \frac12
\, ( \M_a - \N_a )$, whence the commutation relations read
\[   \{ \Q_a,\Q_b \} = \e_{ab}^{\;\;\;\: c} \, \Q_c ,\qquad\quad
     \{ \Q_a,\P_b \} = 0 ,\qquad\quad
     \{ \P_a,\P_b \} = \e_{ab}^{\;\;\;\: c} \, \P_c .
\]
Alternatively, the commutation relations can be written in
the standard form
\be  \{ \Q_+,\Q_- \} = i \Q_3 , \qquad\quad
     \{ \Q_3,\Q_\pm \} = \mp \, i \Q_\pm , \qquad\qquad\quad
     \Q_\pm := \frac1{\sqrt{2}} \, ( \Q_1 \pm i \Q_2 ),
     \label{a}
\ee
and likewise for $\P$.
Again, the infinitesimal canonical transformations generated by
the \fo\ can be integrated, this time to yield an action of the
group $SO_0(2,2)$ (the connected component of $SO(2,2)$ containing
the identity) on the phase space $\cal P$, acting in its defining
four-dimensional \r\ on the vectors $\xi^{(1)}$ and $\xi^{(2)}$:
\[  (\xi^{(1)},\xi^{(2)}) \stackrel{h \in SO_0(2,2)}{\longrightarrow}
    (h \xi^{(1)}, h \xi^{(2)})  \]
(obviously, the set $\cal S$ is stable under the action of $SO_0(2,2)$
on $\mathbb{R}^8$). In addition, there are the discrete fundamental
``\obs s'' (reflections) $\R_{xy}$, $(x,y) \in \{ (u_1,p_1),(u_2,p_2),
(\pi_1,v_1),$ $(\pi_2,v_2) \}$, which change the sign of the coordinates
$x$ and $y$, leaving all other coordinates invariant. Only two of
them are independent (taking into account the action of $SO_0(2,2)$),
say $\R_1 := \R_{\pi_2 v_2}$ and $\R_2 := \R_{\pi_2 v_2} \circ
\R_{u_1 p_1}$. Together with $\Q_a$ and $\P_a$ they generate an
action of the group $O(2,2)$ on the phase space. The action of the
discrete observables $\R_1$ and $\R_2$ on the continuous observables
is given by
\be  \R_1 : \; \Q_a \longleftrightarrow \P_a ,\qquad\quad
     \R_2 : \; (\Q_3,\P_3) \longleftrightarrow (-\Q_3,-\P_3) ,\quad
               (\Q_\pm,\P_\pm) \longleftrightarrow (\Q_\mp,\P_\mp).
     \label{b}
\ee
Finally, there is a $*$--involution (complex conjugation) on the
classical \aoo\ (\ie, the Poisson algebra generated polynomially
by the \fo\ $\Q_a$ and $\P_a$ that comprises the action of $\R_1$
and $\R_2$), acting on the fundamental observables like
\be  (\Q_3,\Q_\pm,\P_3,\P_\pm,\R_1,\R_2) \longrightarrow
     (\Q_3^*,\Q_\pm^*,\P_3^*,\P_\pm^*,\R_1^*,\R_2^*) =
     (\Q_3,\Q_\mp,\P_3,\P_\mp,\R_1,\R_2).
     \label{c}
\ee

\subsection{The \occ}

By definition \cite{Tr}, the \occ\ comprises the conditions which are
imposed on the (Casimir) invariants of the \aoo\ by the ``vanishing''
of the \c s (via functional dependencies between the invariants
of the \aoo\ and the generalized Casimir elements of the \c s \cite{Tr}). 
Upon quantization, these conditions are turned into \r\ conditions
which select the physical \r (s) of the \aoo.

The Lie algebra $so(2,2)$ possesses two quadratic Casimir
``operators'', $\Q^2 := g^{ab} \Q_a \Q_b = \Q_+\Q_- + \Q_-
\Q_+ - \Q_3^2$ and $\P^2 := g^{ab} \P_a \P_b$. If $\Q^2\le
0$ ($\P^2 \le 0$), the sign of $\Q_3$ ($\P_3$) is another
invariant (where the sign can be equal to either plus one,
zero, or minus one).

In the realization of the Lie algebra $so(2,2)$ by the \obs s of
the model at hand, there are two identities which reflect the
functional dependencies between the Casimirs of the \aoo\ and
the Casimir $\L^2 := g^{ab} \L_a \L_b$ of the \c\ algebra:
\be  \Q^2 - \L^2 \equiv 0 ,\qquad\qquad\qquad \P^2 - \L^2 \equiv 0.
     \label{d}
\ee
Upon the vanishing of the \c s they induce the identities
\be  \Q^2 = 0 ,\qquad\qquad\qquad  \P^2 = 0.
     \label{e} \ee
In addition, there are further identities which determine
the signs of $\Q_3$ and $\P_3$. These identities involve the
generators $\Q_a \P_b$ of a Poisson ideal in the \aoo. Only
one of them is independent:
\be
       2 \, \Q_3 \P_3 - [ (\u^2 - \v^2) \, \L_1 - (\u\cdot\p +
       \v\cdot\pe) \, \L_2 + (\u^2 + \v^2) \, \L_3 ] \equiv 0,
       \label{f}
\ee
in the sense that the others can be obtained from it via (repeated)
Poisson bracket operation with $\Q_a$ and $\P_a$.
The vanishing of the \c s induces the identities
\be  \Q_a \P_b = 0,
     \label{k}
\ee
which possess the solutions ($\vee$ is the logical ``or'')
\[   [ \, \Q_a = 0 \quad \forall a \, ] \;\vee\;
     [ \, \P_a = 0 \quad \forall a \, ].
\]
These identities can be translated into conditions involving the
signs of $\Q_3$ and $\P_3$:
\be  [ \, \mbox{sign}(\Q_3) = 0 \, ] \;\vee\;
     [ \, \mbox{sign}(\P_3) = 0 \, ].  \label{cond}
\ee
Now, the \obs s $\O_{ij} = \O_{ij}(\u,\v,\p,\pe)$, regarded as
functions on $\mathbb{R}^8$, vanish for all pairs $(ij)$ of the
indices if and only if $(\u,\v,\p,\pe) \in {\cal S}$. Therefore,
on the phase space $\cal P$ the conditions (\ref{cond}) take on
the form
\be  ( \, [ \, \mbox{sign}(\Q_3) = 0 \, ]  \;\vee\;
          [ \, \mbox{sign}(\P_3) = 0 \, ] \, ) \;\neg\;
     ( \, [ \, \mbox{sign}(\Q_3) = 0 \, ]  \;\wedge\;
          [ \, \mbox{sign}(\P_3) = 0 \, ] \, )
     \label{csig}  \ee
($\wedge$ is ``and'', $\neg$ is ``not'').
The latter conditions, together with the induced identities (\ref{e}),
express the \occ.

\subsection{The structure of the space of physical states}

The topological structure of the space of physical states (\ie,
the reduced phase space) is determined entirely by the \occ. As
the action of the group $SO_0(2,2)$ on (the linear span of) the
\obs s $\Q_a$ and $\P_a$ coincides with the coadjoint action of
$SO_0(2,2)$ on the dual $so^*(2,2) = so^*(2,1)_Q \times so^*(2,
1)_P$ of its Lie algebra $so(2,2) = so(2,1)_Q \times so(2,1)_P$,
the space of physical states is the disjoint union of several
coadjoint orbits for $SO_0(2,2)$ which, in their turn, are labelled
by the values of the invariants.

The identity $\Q^2 = 0$ characterizes the ``light cone'' $\C
_Q$ in $so^*(2,1)_Q$, which falls into three pieces: the
``forward light cone'' $\C^+_Q$ (sign$(\Q_3) = +1$), the ``%
backward light cone'' $\C^-_Q$ (sign$(\Q_3) = -1$), and the
origin $\C^0_Q$ (sign$(\Q_3) = 0$) (and similarly for $\P$).
Therefore, according to the conditions (\ref{e}) and (\ref{csig}),
the space of physical states consists of four pieces (orbits
for $SO_0(2,2)$):
\[   \C_I     = \C^+_Q \times \C^0_P ,\qquad
     \C_{II}  = \C^-_Q \times \C^0_P ,\qquad
     \C_{III} = \C^0_Q \times \C^+_P ,\qquad
     \C_{IV}  = \C^0_Q \times \C^-_P .
\]
These four pieces constitute one orbit $\C = \C_I \cup \C_{II}
\cup \C_{III} \cup \C_{IV}$ for $O(2,2)$, consisting of four
connected components that are mapped onto one another by the
discrete \obs s $\R_1$, $\R_2$ and $\R_1 \circ \R_2$ (corres%
ponding to the four connected components of the group $O(2,2
)$).

\section{Construction of the quantum theory}

I will now construct the quantum theory of the model, following
the quantization scheme outlined in Ref.\ \cite{Tr}.

\subsection{The quantum \aoo}

The first step is the construction of the quantum \aoo. The
quantum \aoo\ is generated polynomially by the \fo\ $\{ \QH_a,
\PH_a \}$, the unambiguous quantum analogs of the classical
\fo\ $\{ \Q_a,\P_a \}$, and an action of the discrete \obs s
$\{ \RH_1,\RH_2 \}$, the quantum analogs of the classical
\obs s $\{ \R_1,\R_2 \}$, has to be defined on it. Up to
quantum corrections, the commutation relations of the \obs s
$\QH_a$ and $\PH_a$ have to be inferred from the Poisson
commutation relations of the corresponding classical \obs s.

The only quantum correction of the classical $so(2,2)$ commutation
relations compatible with the principles formulated in Ref.\ \cite{Tr}
would be a central extension of the algebra $so(2,2)$. However, as
the second cohomology of the algebra $so(2,2)$ is trivial ($so(2,2)$
being semi--simple), there is no non--trivial central extension
available ({\it cf} \cite{Wo}), and the commutation relations have
to be taken to be
\[   [ \QH_a,\QH_b ] = i \hbar \, \e_{ab}^{\;\;\;\: c} \, \QH_c ,\qquad\quad
     [ \QH_a,\PH_b ] = 0 ,\qquad\quad
     [ \PH_a,\PH_b ] = i \hbar \, \e_{ab}^{\;\;\;\: c} \, \PH_c .
\]
The action of the discrete \obs s cannot pick up quantum
corrections. According to equation (\ref{b}), it is given
by ($\QH_\pm$ and $\PH_\pm$ are defined as in the classical
theory, {\it cf}\/ (\ref{a}))
\be  \RH_1 \, (\QH_a,\PH_a) \, \RH_1 = (\PH_a,\QH_a) ,\quad
     \RH_2 \, (\QH_3,\PH_3) \, \RH_2 = (-\QH_3,-\PH_3) ,\quad
     \RH_2 \, (\QH_\pm,\PH_\pm) \, \RH_2 = (\QH_\mp,\PH_\mp) .
     \label{qdis}
\ee

\subsection{The \occ}

The next step is the determination of the form that the conditions
(\ref{e}) and (\ref{csig}) assume upon quantization. The Casimir
operators of $so(2,2)$ being unambiguously defined, $\QH^2 :=
g^{ab} \QH_a \QH_b = \QH_+\QH_- + \QH_-\QH_+ - \QH_3^2$, $\PH
^2 := g^{ab} \PH_a \PH_b$, the only consistent quantum corrections
of the identities (\ref{e}) are by constants:
\[   \QH^2 - c_Q \hbar^2 = 0 ,\qquad\quad  \PH^2 - c_P \hbar^2 = 0,
     \qquad\qquad\quad  c_Q = \mbox{const.},\; c_P = \mbox{const.}.
\]
However, these corrections have to vanish as can be seen by the following
argument: The conditions (\ref{csig}), concerning the signs of $\Q_3$
and $\P_3$, cannot acquire quantum corrections, \ie\ their quantum
counterparts have to read
\be  ( \, [ \, \mbox{sign}(\QH_3) = 0 \, ]  \;\vee\;
          [ \, \mbox{sign}(\PH_3) = 0 \, ] \, ) \;\neg\;
     ( \, [ \, \mbox{sign}(\QH_3) = 0 \, ]  \;\wedge\;
          [ \, \mbox{sign}(\PH_3) = 0 \, ] \, )
     \label{qsig}  \ee
(sign$(\OH) = -1,0,+1$ means that the operator $\OH$ is negative
definite, zero, or positive definite, respectively). But the condition
sign$(\QH_3) = 0$ is only compatible with the identity $\QH^2 = 0$,
and analogously for $\PH^2$, thereby enforcing the identities
\be  \QH^2 = 0 ,\qquad\qquad  \PH^2 = 0 .
     \label{qid}
\ee

\subsection{The physical \r s of the \aoo  \label{3.3}}

The final step is the identification of the physical \r s of
the \aoo, making use of the \occ\ as it is expressed by the
conditions (\ref{qsig}) and (\ref{qid}), and of the hermiticity
relations
\[   (\QH_3,\QH_\pm,\PH_3,\PH_\pm,\RH_1,\RH_2) \longrightarrow
     (\QH_3^\dagger,\QH_\pm^\dagger,\PH_3^\dagger,\PH_\pm^\dagger,
      \RH_1^\dagger,\RH_2^\dagger) =
     (\QH_3,\QH_\mp,\PH_3,\PH_\mp,\RH_1,\RH_2),
\]
which implement the classical $*$--relations (\ref{c}) into the
quantum theory and require the \r s to be hermitian (corresponding
to unitary \r s of the group $O(2,2)$). As the hermitian \r s of
the Lie algebra $so(2,2)$ are the tensor products of the hermitian
\r s of the factors $so(2,1)_Q$ and $so(2,1)_P$, it is sufficient
to determine the latter.

There are three hermitian \ir s of $so(2,1)_Q$ which are selected
uniquely by the conditions $\QH^2 = 0$ and sign$(\QH_3) = -1, \,
0, \, +1$ (for a classification of the hermitian \ir s of $so(2,1)$
see Ref.\ \cite{Wy}): the \r\ $D^-_Q(-1)$, the trivial \r\ $\id_Q$,
and the \r\ $D^+_Q(-1)$, respectively (in the notation of Ref.\
\cite{Wy}). The \r s are spanned by orthonormal states ($m \in
\mathbb{N}$)
\[  | -m \rangle_Q \quad \mbox{for} \quad D^-_Q(-1) ,\qquad\quad
    | 0 \rangle_Q \quad  \mbox{for} \quad \id_Q ,\qquad\quad
    | m \rangle_Q \quad  \mbox{for} \quad D^+_Q(-1) .\]
The action of $\QH_3$ and $\QH_\pm$ on these states is given by
($n = -m,0,m$)
\be  \QH_3 \, | n \rangle_Q = \hbar n \, | n \rangle_Q ,\qquad\quad
     \QH_\pm \, | n \rangle_Q = \frac\hbar{\sqrt{2}} \, \sqrt{n(n\pm1)}
     \: | n \pm 1 \rangle_Q .
     \label{g}
\ee
Analogously, there are three hermitian \ir s $D^-_P(-1)$, $\id_P$,
$D^+_P(-1)$ of $so(2,1)_P$, selected uniquely by the conditions
$\PH^2 = 0$ and sign$(\PH_3) = -1, \, 0, \, +1$, respectively.

As a consequence, the \occ, conditions (\ref{qsig}) and (\ref{qid}),
selects uniquely four hermitian \ir s of the Lie algebra $so(2,2)$,
corresponding to the four orbits which make up the classical space
of physical states, namely $D_I = D^+_Q(-1) \otimes \id_P$, $D_{II}
= D^-_Q(-1) \otimes \id_P$, $D_{III} = \id_Q \otimes D^+_P(-1)$,
$D_{IV} = \id_Q \otimes D^-_P(-1)$,
spanned by the states $| m,0 \rangle = | m \rangle_Q \otimes | 0
\rangle_P$, $| -m,0 \rangle = | -m \rangle_Q \otimes | 0 \rangle_P$,
$| 0,m \rangle = | 0 \rangle_Q \otimes | m \rangle_P$,
$| 0,-m \rangle = | 0 \rangle_Q \otimes | -m \rangle_P$,
respectively. The physical Hilbert space $\HI_{phys}$ is the direct
sum of (the carrier spaces of) these four \r s.

Finally, defining the action of the discrete \obs s $\RH_1$ and
$\RH_2$ on $\HI_{phys}$ by
\[  \RH_1 \, | n,0 \rangle = | 0,n \rangle ,\qquad
    \RH_1 \, | 0,n \rangle = | n,0 \rangle ,\qquad
    \RH_2 \, | n,0 \rangle = | -n,0 \rangle ,\qquad
    \RH_2 \, | 0,n \rangle = | 0,-n \rangle ,
\]
the requirements (\ref{qdis}) are satisfied, and $\HI_{phys}$
carries one unitary \ir\ $D = D_I \oplus D_{II} \oplus D_{III}
\oplus D_{IV}$ of the group $O(2,2)$.

\section{Comparison with the Dirac quantization}

The phase space $\cal P$ not being a cotangent bundle, the canonical
Dirac quantization scheme does not seem to be an appropriate setting
for the construction of the quantum theory of the $\SL$ model.
Nevertheless, it is possible to - at least abstractly - complete the
Dirac quantization that has been left unfinished in Ref.\ \cite{MRT}.

\subsection{Constraints and physical states}

The authors of Ref.\ \cite{MRT} represent the \c s as symmetric operators
$\HH_1$ ($=: \LH_1 + \LH_3$), $\HH_2$ ($=: \LH_1 - \LH_3$) and $\DH$
($=: 2 \,\LH_2$) with the $so(2,1)$ commutation relations $[ \LH_a,\LH_b ]
= i \hbar \, \e_{ab}^{\;\;\;\: c} \, \LH_c$ on the Hilbert space $\HI
= L^2(\mathbb{R}^4,{\rm d}^2u \, {\rm d}^2v)$. The concrete expressions
for these operators can be obtained from the classical ones upon
substituting the multiplicative operators $\hat{\u} = \u$ and $\hat
{\v} = \v$ for the classical coordinates $\u$ and $\v$, and the differential
operators $\hat{\p} = -i\hbar \, \vec{\nabla}_u$ and $\hat{\pe} = -i\hbar
\, \vec{\nabla}_v$ for the classical momenta $\p$ and $\pe$,
leaving the order of the coordinates and momenta unchanged.
They determine the ``physical states'' of the system as distributional
solutions of the differential equations $\LH_a \, \Psi(\u,\v) = 0$. In
polar coordinates, $\u = ( u \,\cos\alpha, u\,\sin\alpha )^T$, $\v = (v
\,\cos\beta, v\,\sin\beta )^T,$ these solutions are given by
\[   \Psi_{m,\epsilon} (u,v,\alpha,\beta) = e^{im (\alpha +
     \epsilon \beta)} \: J_m (uv/\hbar) ,
\]
where $\epsilon = \pm 1$, $m \in \mathbb{Z}$, and $J_m(z)$ are
Bessel functions. (By the way: there are also other ``physical
states'', {\it e.g.}\ the states $\Psi_\pm = \exp\{ \pm i \,\u
\cdot \v / \hbar \}$ and all states that can be obtained from
them via the action of \obs s.) The authors do not specify an
inner product on these states, \ie\ they do not construct a physical
Hilbert space. Furthermore, they represent only the generators
of the maximal compact subgroup $SO(2) \times SO(2)$ of $SO_0(
2,2)$ as symmetric operators on $\HI$, which is not sufficient
for the determination of the \r s of the full \aoo.

\subsection{Observables}

The classical \obs s $\Q_a$ and $\P_a$ can be represented as symmetric
operators $\QH_a$ and $\PH_a$ on $\HI$ in the same way as the \c s
(\ie, by the same substitutions; there are no ordering ambiguities).
The \obs s obey the commutation relations
\[   [ \QH_a,\QH_b ] = i \hbar \, \e_{ab}^{\;\;\;\: c} \, \QH_c ,\qquad\quad
     [ \QH_a,\PH_b ] = 0 ,\qquad\quad
     [ \PH_a,\PH_b ] = i \hbar \, \e_{ab}^{\;\;\;\: c} \, \PH_c \]
and commute with the \c\ operators. With the concrete expressions
for the \obs s and \c s the quantum counterparts of the classical
identities (\ref{d}) and (\ref{f}) can be computed explicitly. They
read
\be  \QH^2 - \LH^2 \equiv 0 ,\qquad\qquad\qquad \PH^2 - \LH^2 \equiv 0
     \label{h}
\ee
and
\be
       2 \, \QH_3 \PH_3 - [ (\hat{\u}^2 - \hat{\v}^2) \, \LH_1 -
       (\hat{\u} \cdot \hat{\p} + \hat{\v} \cdot \hat{\pe}) \, \LH_2 +
       (\hat{\u}^2 + \hat{\v}^2) \, \LH_3 ] \equiv 0,
       \label{i}
\ee
Again, the missing identities can be obtained from (\ref{i})
via (repeated) commutator operation with $\QH_a$ and $\PH_a$.
On the linear space $\V$ of ``physical states'' the identities
(\ref{h}) and (\ref{i}) induce the identities
\[   \QH^2 = 0 ,\qquad\quad  \PH^2 = 0 ,\qquad\qquad
     \QH_a \PH_b = 0,
\]
which coincide with the classical identities (\ref{e}) and
(\ref{k}).

\subsection{Representations}

The \r s of $so(2,1)_Q \times so(2,1)_P$ which are realized on
the linear space $\V$ of ``physical states'' can be identified
by explicitly computing the action of $\QH_3$, $\QH_\pm$, $\PH
_3$ and $\PH_\pm$ on the states $\Psi_{m,\epsilon}$. In polar
coordinates these operators are given by ($\OH_a(+1) := \QH_a$,
$\OH_a(-1) := \PH_a$)
\[  \OH_3(\epsilon) = - \frac{i\hbar}2 \, (\pa_\alpha + \epsilon\pa_\beta) ,
    \qquad
    \OH_\pm(\epsilon) = \frac\hbar{\sqrt{8}} \,
               e^{\pm i (\alpha + \epsilon \beta)} \:
               \Big[ \frac{\epsilon}{v} \, \pa_u \pa_\beta +
               \frac1{u} \, \pa_v \pa_\alpha \mp
               i \, \Big( uv + \pa_u \pa_v - \frac{\epsilon}{uv} \,
               \pa_\alpha \pa_\beta \Big) \Big] .
\]
The action of the operators $\QH_a$ on the states $\Psi_{m,-}$
is trivial: $\QH_a \, \Psi_{m,-} = 0$. On the states $\Psi_{m,
+}$ $\QH_3$ and $\QH_\pm$ act as
\be  \QH_3 \, \Psi_{m,+} = \hbar m \, \Psi_{m,+} ,\qquad\quad
     \QH_\pm \, \Psi_{m,+} = \frac{\mp i}{\sqrt{2}} \, \hbar m \,
     \Psi_{m \pm 1,+}
     \label{j}
\ee
(and similarly for $\PH_3$ and $\PH_\pm$, with the r\^{o}les of
$\Psi_{m,+}$ and $\Psi_{m,-}$ interchanged). Equation (\ref{j})
(and its counterpart for $\PH_\pm$) shows that the linear space
$\V$ contains five invariant subspaces for the action of $so(2,
2)$: $\V_Q^+ = \{ \Psi_{m,+} | m \ge 0 \}$, $\V_Q^- = \{ \Psi_{
m,+} | m \le 0 \}$, $\V_P^+ = \{ \Psi_{m,-} | m \ge 0 \}$, $\V_
P^- = \{ \Psi_{m,-} | m \le 0 \}$, and $\V_0 = \{ \Psi_{0,+} =
\Psi_{0,-} \}$. The space $\V$ is not the direct sum of these
invariant subspaces, as all spaces have the subspace $\V_0$ in
common. This means, that the \r\ of $so(2,2)$ on $\V$ is reducible,
but not fully reducible (\ie, it is indecomposable).

The invariant subspace $\V_Q^+$ carries a \r\ of $so(2,1)_Q$.
In the notation of Ref.\ \cite{Wy} it is the \r\ $D^+_Q(0)$. This
\r\ is also indecomposable, as it contains the invariant sub%
space $\V_0$ and as the action of $\QH_-$ maps $\Psi_{1,+}$
into $\V_0$. This latter fact means that the sign of $\QH_3$
is not invariant (it changes from plus one to zero), showing
that the \occ\ is not reproduced correctly. Moreover, the \r\
is not unitary. For, assume that there exists a scalar product
$\langle\cdot|\cdot\rangle$ on $\V_Q^+$, such that $\langle
\Phi_{m,+} | \Phi_{m^\prime,+} \rangle = \delta_{m m^\prime}$
($m,m^\prime \ge 0$, $\Phi_{m,+} = {\cal N}_m \, \Psi_{m,+}$,
${\cal N}_m$ non-zero and finite normalization constants) and
such that $\QH_+^\dagger = \QH_-$. Then, from equation (\ref{j})
one finds
\[  \QH_- \, \Phi_{1,+} = \frac{i\hbar}{\sqrt{2}} \, \frac{{\cal N}_1}
                          {{\cal N}_0} \; \Phi_{0,+} ,\qquad\quad
    \QH_+ \QH_- \, \Phi_{1,+} = 0 , \]
which implies the contradiction
\[  0 = \langle \Phi_{1,+} | \QH_+ \QH_- | \Phi_{1,+} \rangle
      = \langle \QH_- \, \Phi_{1,+} | \QH_- \, \Phi_{1,+} \rangle
      = \frac{\hbar^2}2 \, \Big| \frac{{\cal N}_1}{{\cal N}_0} \Big|^2
      \neq 0 . \]

However, it is possible to (abstractly) introduce a degenerate scalar
product on (an abstract linear space that is isomorphic to) $\V_Q^+$
which allows to turn $\V_Q^+$ into a Hilbert space that carries the
\ir\ $D^+_Q(-1)$. For this purpose define states $| m )$,
$m \in \mathbb{N}_0$, and abstract operators $\QH_3$ and $\QH_\pm$
which act on these states as in equation (\ref{j}), \ie\ as
\[   \QH_3 \, | m ) = \hbar m \, | m ) ,\qquad\quad
     \QH_\pm \, | m ) = \frac{\mp i}{\sqrt{2}} \, \hbar m \,
     | m \pm 1 ) .
\]
Introduce a degenerate scalar product $(\cdot|\cdot)$ on these
states by requiring
\[  (0|0) = 0 ,\qquad\quad (0|m) = 0 ,\qquad\quad
    (m|m^\prime) = |N_m|^{-2} \, \delta_{m m^\prime}  \]
($m,m^\prime \in \mathbb{N}$, $N_m$ are non-zero and finite
complex normalization constants). Denote the equivalence
classes (\ie, states modulo zero norm states) of $N_m \,
| m )$, $m \in \mathbb{N}$, by $| m \rangle$. The induced
scalar product on these states is $\langle m | m^\prime
\rangle = \delta_{m m^\prime}$, and the action of the operators
$\QH_3$ and $\QH_\pm$ is given by
\[   \QH_3 \, | m \rangle = \hbar m \, | m \rangle ,\qquad\quad
     \QH_\pm \, | m \rangle = \frac{\mp i}{\sqrt{2}} \, \hbar m \,
     \frac{N_m}{N_{m \pm 1}} \: | m \pm 1 \rangle .
\]
The normalization constants $N_m$ can be fixed (up to a phase)
by requiring that $\QH_+^\dagger = \QH_-$. This yields the
condition $| N_m / N_{m+1} |^2 = (m+1)/m$, which can be satisfied
by putting (fixing also the phase)
\[   N_m = \frac{i^{-m}}{\sqrt{m}} . \]
As a consequence, the operators $\QH_\pm$ act on the states $| m
\rangle$ as
\[  \QH_\pm \, | m \rangle = \frac\hbar{\sqrt{2}} \, \sqrt{m(m\pm1)}
    \: | m \pm 1 \rangle , \]
and a comparison with equation (\ref{g}) reveals, that the Hilbert
space $\HI^+_Q$, which is spanned by the states $| m \rangle$ and
equipped with the scalar product $\langle\cdot|\cdot\rangle$,
carries the hermitian \ir\ $D^+_Q(-1)$ of $so(2,1)_Q$ (or the
\r\ $D^+_Q(-1) \otimes \id_P$ of $so(2,1)_Q \times so(2,1)_P$,
\ie, the states $|m\rangle$ can be identified with the states
$|m,0\rangle$ of Sec.\ \ref{3.3}).

In an analogous fashion, the invariant subspaces $\V_Q^-$,
$\V_P^+$ and $\V_P^-$ can be turned into Hilbert spaces $
\HI^-_Q$, $\HI^+_P$ and $\HI^-_P$ which carry the \ir s
$D^-_Q(-1) \otimes \id_P$, $\id_Q \otimes D^+_P(-1)$ and
$\id_Q \otimes D^-_P(-1)$, respectively, thereby establishing
a one-to-one correspondence between the basis states of the
space $\V - \V_0$ and those of the direct sum of the Hilbert
spaces $\HI^\pm_Q$ and $\HI^\pm_P$. In this way the results
of Sec.\ \ref{3.3} can be reproduced.

\section{Conclusions}

In this note I hope to have demonstrated that the algebraic method
is an effective and natural (in the sense of being well adapted to
the problem) tool for the construction of the quantum theory of
constrained systems.

Let me point out a peculiar structural feature that makes the $\SL$
model an interesting example for the application of
the method, emphasizing the crucial r\^{o}le that is played by the
correct identification of the \occ: For the determination of the
\occ\ it is necessary to take into
account not only the direct functional dependencies (equations
(\ref{d})) between the Casimirs of the \aoo\ and those of the \c\
algebra, but also dependencies (equation (\ref{f}) and its Poisson
bracket transforms) between the
generators of an ideal within the \aoo\ and generalized Casimir
elements of the \c s. However, the conditions that are imposed by
the vanishing of the \c s via the latter dependencies can be
reformulated as restrictions on the values of the invariants of
the \aoo, {\it cf}\/ equation (\ref{csig}).

\vspace{5mm}
\noindent
{\bf Acknowledgement}: I thank G.\ Handrich for drawing my attention
to the peculiar structure of the $\SL$ model, for several helpful
discussions, and for critically reading the manuscript.

\vspace{5mm}
\noindent
{\bf Postscript}: Some time after the completion of the present
note the preprint \cite{LR} appeared, where the authors Louko
and Rovelli construct the quantum theory of the $\SL$ model
using the quantization schemes of ``Algebraic Quantization''
and ``Refined Algebraic Quantization'', arriving at essentially
the same results as in the treatment given here. A comparison of
the conceptual and technical advantages and disadvantages
of the various schemes is left to the reader. I just want to point
out that the approach presented here does not make use of any
``input motivated by the structure of the classical \c s'' \cite{LR},
and that it does not depend on ``making successful choices in the
`early' steps'' which ``may require hindsight from the `later'
steps'' \cite{LR}. To be sure, just like ``Algebraic Quantization''
and ``Refined Algebraic Quantization'' the algebraic quantization
scheme followed here is not a ``{\it prescription}
for quantization'' \cite{LR} which can do without any input, but
here the input is taken exclusively from the \obs\ sector of the
system in question -- in the form of correspondence and consistency
requirements which are imposed on {\it \obs} quantities.

\end{document}